\documentclass[aps,showpacs,preprint,amsmath,amssymb]{revtex4}
\usepackage{graphics}

\begin{document}
\title{Exact calculation of the skyrmion lifetime in a ferromagnetic Bose condensate}
\author{Yunbo Zhang$^{1,2}$, Wei-Dong Li$^2$, Lu Li$^2$, H.J.W. M\"{u}ller-Kirsten$^1 $}
\affiliation{$^1$Department of Physics, University of Kaiserslautern, D-67653, Kaiserslautern, Germany \\
$^2$Department of Physics and Institute of Theoretical Physics, Shanxi University, Taiyuan 030006, P.R. China}
\date{\today}

\begin{abstract}
The tunneling rate of a skyrmion in ferromagnetic spin-1/2 Bose condensates
through an off-centered potential barrier is calculated exactly with the
periodic instanton method. The prefactor is shown to depend on the chemical
potential of the core atoms, at which level the atom tunnels. Our results
can be readily extended to estimate the lifetime of other
topological excitations in the condensate, such as vortices and monopoles.
\end{abstract}
\maketitle

\section{Introduction}

Macroscopic quantum tunneling(MQT), the tunneling of a macroscopic variable
of a macroscopic system, has recently received much attention in studies
of Bose-Einstein condensation(BEC). The tunneling of a condensate through an
optical lattice potential \cite{Anderson,Cataliotti} provides an atomic
physics analogue of a Josephson junction array, while in principle the
analogue of a single junction can be realized by two condensates confined in
a double well potential\cite{Andrews,Smerzi}. The recent experimental
success in all-optical trapping of an atomic condensate \cite{optical} opens
the prospect of studies into the internal structure of spinor BECs,
including the possibility of creating some topological excitations \cite{top} such as
skyrmions, monopoles, merons or axis-symmetric or non axis-symmetric
vortices both for antiferromagnetic and ferromagnetic condensates. Among
various topological structures, the Mermin-Ho (MH)\cite{MH} and
Anderson-Toulouse (AT)\cite{AT} coreless non-singular vortices are
demonstrated to be thermodynamically stable in ferromagnetic spinor
Bose-Einstein condensates with the hyperfine state $F=1$\cite{Mizushima}.
Skyrmions, which do not have an ordinary vortex core due to the spin degree
of freedom, are also proposed in the spinor BEC \cite{RA,BCS,MZS}and are shown not to be
thermodynamically stable objects without rotation\cite{Usama1,Usama2}. Once
created, the radius of such a skyrmion shrinks to zero so that one must
detect and manipulate it in the duration of its lifetime.

The skyrmion texture in a ferromagnetic spinor condensate can be described
conveniently by a position-dependent spinor\cite{Usama2} 
\begin{equation}
\zeta ({\bf r})=\exp \left\{ -\frac iS\frac{\omega (r)}r{\bf r}\cdot {\bf S}%
\right\} \zeta ^Z.
\end{equation}
The constant spinor $\zeta ^Z$ is the usual basis that diagonalizes the $S_z$
component of the spin matrices ${\bf S}$ and $\omega (r)$ is a real function
of radius $r$ satisfying the boundary conditions $\omega (0)=2\pi $ and $%
\lim_{r\rightarrow \infty }\omega (r)=0$. For the skyrmion with size of the
order of the correlation length $\xi =1/\sqrt{8\pi an_\infty }$ or less,
where $n_\infty $ is the average atomic density and $a$ the $s$-wave
interatomic scattering length, the problem can be reduced to a nonlinear
Schr\"{o}dinger equation by an ansatz for $\omega (r)$ with the gradient
term $\left| \nabla \zeta ({\bf r})\right| ^2$ in the Gross-Pitaevskii
energy functional regarded as some external potential $V(r)=\hbar ^2\left|
\nabla \zeta ({\bf r})\right| ^2/2m$. In the spin-1/2 case, for example, the
latter takes the form\cite{Usama1} 
\begin{equation}
V(r)=\frac{\hbar ^2}{2m}\frac{32}{\lambda ^2}\frac{\left( r/\lambda \right)
^2\left[ 3+2\left( r/\lambda \right) ^4+3\left( r/\lambda \right) ^8\right] 
}{\left[ 1+\left( r/\lambda \right) ^4\right] ^4}
\end{equation}
for an ansatz $\omega (r)=4\cot ^{-1}[\left( r/\lambda \right) ^2]$, where
the variational parameter $\lambda $ corresponds physically to the size of
the skyrmion. The lifetime of the skyrmion is estimated by employing a WKB
expression for the tunneling rate $\Gamma =\frac{\omega _0}{2\pi }%
e^{-S_c/\hbar }$, with $S_c$ the action through the barrier and $\omega _0$
the characteristic frequency of the harmonic potential which was used to
approximate the potential $V(r)$. Due to the inaccuracy of the prefactor in
the tunneling rate, which makes it difficult to give a reliable result for
the decay rate, more efficient methods are needed for the investigation of
this problem.

The instanton method as a powerful tool for dealing with quantum tunneling
phenomena has generally been used in the evaluation of the splitting of
degenerate ground states or the escape rate from metastable ground states
\cite{Coleman}. A method of evaluating quantum mechanical tunneling at 
excited energy states has been developed recently by means of periodic 
instantons and bounces \cite{Liang1},
which are characterized by nonzero energy and satisfy manifestly nonvacuum
boundary conditions. Solvable models include, level splittings for the
double well and sinh-Gordon potentials, decay rates for the inverted double
well and cubic potentials, and energy band structures of the sine-Gordon and
Lam\'{e} potentials \cite{Liang1,models}. The off-centered potential barriers serve as another
class of physical systems which permit analytical evaluation.

In this paper, we investigate the tunneling behavior of the skyrmion from the
core to the outer region through an off-centered barrier. We first solve the
equation of motion in the Euclidean version to find the classical
configuration, which in our case is a bounce. In Section III we present the
formalism of the periodic instanton theory for tunneling and calculate the
decay rate exactly. The results obtained are applied to estimate
the rate of shrinking of the skyrmion in the two-component ferromagnetic
Bose-Einstein condensate. Finally we summarize the main results.

\section{Instantons for off-centered potential barrier}

In this section we consider the instanton solution for the tunneling in an
off-centered potential barrier as depicted in Figure 1 in which the potential
\begin{equation}
V(r)=\frac{Ar^2}{\left( 1+Br^2\right) ^2}  \label{voc}
\end{equation}
takes into account all the main features of the real barriers in the
skyrmion excitation in the condensate, both in spin-1/2 and spin-1
condensates, for any reasonable ansatz $\omega (r)$. We observe here some
essential conditions for this simplified model: First, as a function the ansatz 
for $\omega $ should decrease monotonically from $2\pi $ to $0$,
since this will correspond to the smallest gradient energy for the spin
deformations; correspondingly, this excludes any oscillation in the decrease of
the potential $V(r)$ when $r$ tends to $+\infty $. Furthermore, $%
V(r)$ should be an off-centered potential barrier with a maximum height $%
V(r_m)$ at $r=r_m,$ and $V(0)=V(+\infty )=0$. Finally, the potential should
be an even function of $r$, and to avoid the point $r=0$ becoming a
singularity, we have $V^{\prime \prime }(0)>0$ so that the harmonic
oscillation frequency $\omega _0$ can be well defined as $\sqrt{V^{\prime
\prime }(0)/m}$. The barrier (\ref{voc}) is just the simplest form
fulfilling the above requirements, with parameter $A$ determining the
barrier height, parameter $B$ the position of the barrier: 
\begin{equation}
r_m=\sqrt{1/B},\qquad V_m=\frac A{4B},\qquad \omega _0=\sqrt{2A/m}.
\label{bh}
\end{equation}

\begin{figure*}
\includegraphics{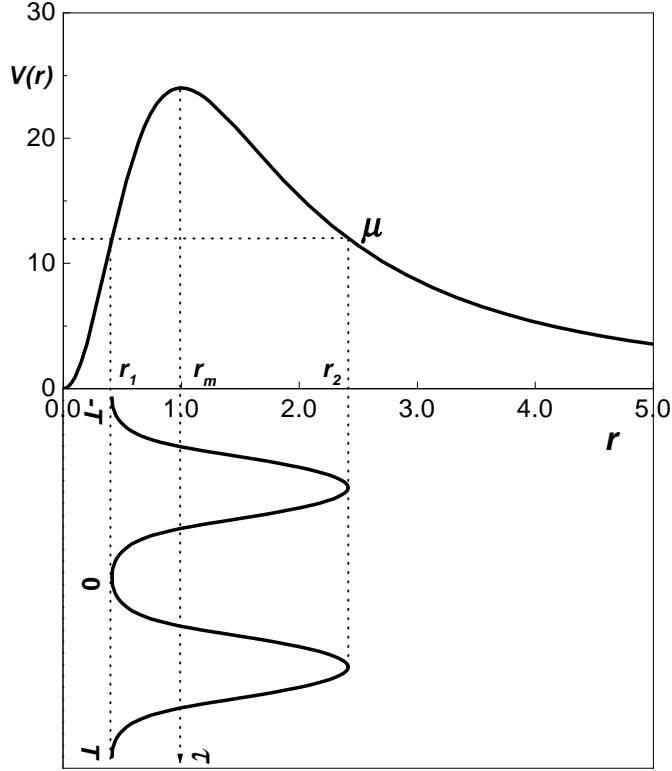}
\caption{The off-centered potential and the bounce configuration in
two imaginary time periods. For the spin-1/2 $^{87}Rb$ condensate
the potential and the radius are in units of $\hbar^2/2 m \xi^2$ and $\xi$, respectively, 
and the parameters are chosen as $A=96$ and $B=1$,
where the size of the skyrmion $\lambda$ used is approximately 
the corelation length $\xi$ corresponding to 20 core atoms.}
\end{figure*}

To estimate the lifetime of the skyrmion, we calculate the tunneling rate
from the core to the outer region through a barrier, the core atoms having a
chemical potential $\mu _{core}$(hereafter abbreviated as $\mu $). The first
step of the instanton method is the so-called Wick rotation of a phase space
corresponding to a transformation to imaginary time $\tau =it$. After the
transformation the Lagrangian is replaced by its Euclidean counterpart 
\begin{equation}
{\cal L}=\frac 12m\left( \frac{dr}{d\tau }\right) ^2+V(r).  \label{lag}
\end{equation}
The classical solution $r_c$ which minimizes the corresponding Euclidean
action satisfies the equation 
\begin{equation}
\frac 12m\left( \frac{dr_c}{d\tau }\right) ^2-V(r)=-E_{cl},  \label{mo}
\end{equation}
which can be viewed as the equation of motion for a particle of mass $m$
with energy $-E_{cl}$ in a potential $-V$. For the tunneling process in the
condensate, we assume that the skyrmion has decreased to a size for which
the barrier is so high that the overlap between the core atoms and the
external atoms is exponentially small. The classical turning points on both
sides of the barrier can be determined by the relation $V(r_{1,2})=\mu $ as
suggested in ref. \cite{Usama1} 
\begin{equation}
r_{1,2}=\frac{\sqrt{A/\mu }}{2B}\left( 1\mp \sqrt{1-4\mu B/A}\right) ,\qquad
0<\mu <V_m.
\end{equation}
The reason why we can handle a nonlinear problem by means of a linear
equation of motion is that we discuss the tunneling behavior in the barrier
region where the nonlinear interaction is negligibly small. Furthermore the
condensate at the ground state can be well described by a macroscopic
wavefunction with unique phase just as in the single particle case. However,
there are obvious differences between the BEC tunneling system and the usual
one-body problem, i.e. the nonlinear interaction contributes a finite
chemical potential $\mu $, which replaces the integration constant $E_{cl}$
on the right hand side of eq.(\ref{mo}).

The classical configuration is a bounce which is the solution of eq.(\ref{mo})
and can be expressed in an implicit form 
\begin{equation}
f(r_c)=\omega _c\tau .  \label{bounce}
\end{equation}
Here we have assigned a characteristic frequency 
\begin{equation}
\omega _c=\sqrt{\frac{2\mu B^2}m},
\end{equation}
and the function $f$ takes the form 
\begin{equation}
f(r_c)=\frac 1{r_2}u_1+Br_2\left[ E(u_1)-k^2snu_1cdu_1\right] 
\end{equation}
where $sn$, $cd$ are two Jacobian elliptic functions, $u_1=F(\varphi ,k)$ and $E(u_1)$
are the first and second kind of incomplete elliptic integrals with modulus $%
k=\sqrt{1-r_1^2/r_2^2}$ respectively\cite{Byrd}, and 
\begin{equation}
\varphi =\sin ^{-1}\sqrt{\frac{r_2^2(r_c^2-r_1^2)}{r_c^2(r_2^2-r_1^2)}}.
\end{equation}
The solution is subject to the following boundary conditions 
\begin{eqnarray}
\tau  &=&0,r=r_1,  \nonumber \\
\tau  &=&\pm T,r=r_1,  \label{boundary} \\
\tau  &=&\pm T/2,r=r_2,  \nonumber
\end{eqnarray}
and exhibits periodic oscillation with imaginary time period 
\begin{equation}
T=\frac 2{\omega _c}\left( \frac 1{r_2}K(k)+Br_2E(k)\right), 
\end{equation}
where $K(k)$ and $E(k)$ are the first and second kind of complete elliptic
integrals with modulus $k$, respectively. In Figure 1 we depict the periodic
oscillation of this pseudoparticle in two periods. A remarkable feature of
this bounce configuration is that there is no vacuum analogue as in the case
of the simple cubic metastable potential, the latter describing the
tunneling behavior of a particle located at the ground state. As the energy $%
E_{cl}$ (or the chemical potential $\mu $) approaches zero, the barrier will
become infinitely thick and the particle confined in the core region will be
stable, with no possibility to tunnel to the outer region.

\section{Exact Calculation of the Decay Rate}

The tunneling rate of the condensate core atoms was given by a simple
expression of the form $\Gamma =Pe^{-W/\hbar }$, where $P$ and $W$ are
coefficients which depend on the detailed form of the metastable potential.
The quantity $W$ appearing in the exponential is the Euclidean action of the
bounce solution and gives the dominant contribution to the tunneling rate,
while the prefactor $P$ originates from the fluctuation around the classical
configuration. For a rather rough estimate, $P$ is often taken to be the
attempt frequency $\omega _0/2\pi $ as was done in ref.\cite{Usama1}. However,
as we will show below, this simple evaluation is not accurate. This paper
provides a powerful instanton tool to obtain this prefactor.

We recall for the sake of convenience the main ideas of the periodic
instanton approach. Let's first denote the wavefunction of the core atom condensate 
with chemical potential $\mu $ by $\zeta ({\bf r}) \mid \psi _\mu >$, where
$\mid \psi _\mu >=\sqrt{n({\bf r})}$ originates from the density and satisfies 
the Gross-Pitaevskii equation for single component
\begin{equation}
H\mid \psi _\mu > =\mu \mid \psi _\mu >, \qquad H=-\frac{\hbar ^2 \nabla ^2}{2 m}
+V({\bf r})+g \mid \psi _\mu \mid ^2
\end{equation}
The effective external potential $V(\bf r)$ \cite{Usama2} comes from the gradient 
term of the spinor $\left| \nabla \zeta ({\bf r})\right| ^2$, and the term with 
coupling constant $g$ represents the strength of the interatomic interactions.

The tunneling effect leads to the decay of the metastable state. In the case
under discussion, the nonconservation of an exponentially small probability current
through the barrier requires that the chemical potential has an imaginary part
proportional to the decay rate \cite{Affleck}, $\Gamma =\frac 2\hbar 
\mathop{\rm Im}%
\mu $. Consider the transition amplitude from the state $\mid \psi _\mu >$
to itself due to quantum tunneling in Euclidean time period $T$. The
amplitude is simply 
\begin{equation}
A=<\psi _\mu |e^{-HT/\hbar }\mid \psi _\mu >=e^{-\mu T/\hbar }.  \label{aa}
\end{equation}
In general the transition amplitude is calculated with the help of the path
integral method as
\begin{equation}
A=\int \psi _\mu ^{*}(r_f)\psi _\mu (r_i){\cal K}(r_f,T;r_i,0)dr_idr_f,
\end{equation}
where $r_f=r_c(T),r_i=r_c(0)$ denote the end points of the bounce motion,
which tend to the turning points $r_1$ (see the boundary condition eq. (\ref
{boundary})). The wave functions $\psi _\mu (r_i),\psi _\mu (r_f)$ in the
barrier region are specified in the WKB approximation as\cite{Landau} 
\begin{eqnarray}
\psi _\mu (r) &=&\frac C{\sqrt{\left| p\right| }}\exp \left( -\frac 1\hbar
\int_{r_1}^rpdr\right) , \\
p &=&\sqrt{2m(\mu -V(r))},
\end{eqnarray}
with $C$ a normalization constant to be determined below. The Feynman kernel
is defined as the summation over all possible classical paths $r(\tau )$%
\begin{equation}
{\cal K}(r_f,T;r_i,0)=\int_{r_i}^{r_f}{\cal D}\{r\}\exp (-S/\hbar ) \label{kernel}.
\end{equation}

We know that the classical solution (\ref{bounce}) which minimizes the
action $S$ gives rise to the major contribution to the above kernel
integral, while the quantum fluctuation around it results in a prefactor $P$%
. In the period $T$ the bounce eq. (\ref{bounce}) completes one oscillation
and crosses the barrier region twice, back and forth. The Euclidean action
is thus calculated in this period as 
\begin{eqnarray}
S_E &=&\int {\cal L}(r,\dot{r})d\tau =\int_0^T\left( m\left( \frac{dr_c}{%
d\tau }\right) ^2+\mu \right) d\tau  \\
&=&W+\mu T,  \nonumber
\end{eqnarray}
while the so-called abbreviated Euclidean action\cite{Weiss} 
\begin{equation}
W=2\int_{r_1}^{r_2}dr\sqrt{2m(V(r)-\mu )}  \label{w1}
\end{equation}
can be expressed in terms of elliptic integrals 
\begin{equation}
W=\frac 4{\omega _c}\left[ \frac{Ar_1^2\Pi (\alpha ^2,k)}{r_2\left(
1+Br_1^2\right) }-\frac \mu {r_2}K(k)-\mu Br_2E(k)\right]   \label{w2}
\end{equation}
with $\Pi (\alpha ^2,k)$ the complete elliptic integral with the parameter 
\begin{equation}
\alpha ^2=\frac{k^2}{1+Br_1^2}.
\end{equation}
It is obvious from the potential that $A,B>0,$ so $0<\alpha ^2<k^2$, the
third elliptic integral is complete and belongs to the case III\cite{Byrd}.

\begin{figure*}
\includegraphics{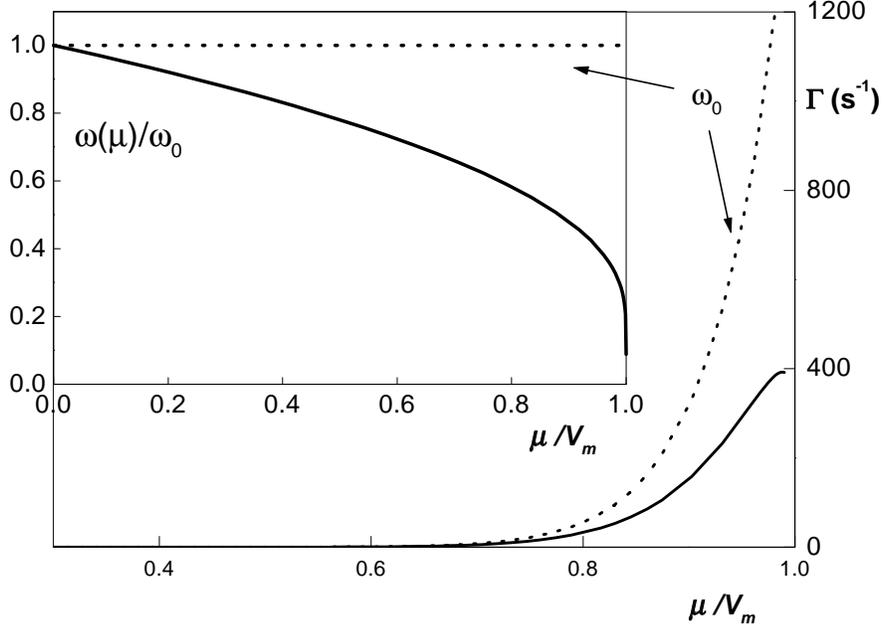}
\caption{The decay rate as a function of the chemical potential.
Inset: the chemical potential dependent frequency $\omega $. The solid
curves represent our exact result while the dotted curves correspond to the
case for constant attempt frequency $\omega _0.$ All curves are calculated
for the parameter of a $^{87}Rb$ spinor condensate and the chemical
potential $\mu $ is given in units of the barrier height $V_m$.}
\end{figure*}

The imaginary part of the chemical potential can be derived by considering
the amplitude $A$ as the sum of contributions from any number of bounces\cite
{Liang1}. The zero bounce contribution results in the real part of the
chemical potential 
\begin{equation}
A^{(0)}=e^{-\mu T/\hbar }.
\end{equation}
The one bounce contribution comes from the classical configuration with
period $T$, and can be obtained by expanding the kernel (\ref{kernel}) 
around the bounce (\ref{bounce}) 
\begin{equation}
A^{(1)}=-iT\frac{C^2}me^{-W/\hbar }e^{-\mu T/\hbar }.
\end{equation}
Generalizing to the case of $n$ bounces straightforwardly, i.e., assuming
the pseudoparticle completing $n$ oscillations in the period $T$, one has 
\begin{equation}
A^{(n)}=\left( -i\right) ^n\frac{T^n}{n!}\left( \frac{C^2}m\right)
^ne^{-nW/\hbar }e^{-\mu T/\hbar }.
\end{equation}
The total transition amplitude is given by the sum over all bounce
contributions 
\begin{equation}
A=\sum_nA^{(n)}=e^{-\mu T/\hbar }\exp \left( -iT\frac{C^2}me^{-W/\hbar
}\right).   \label{a}
\end{equation}
The imaginary part of the chemical potential is obtained by comparing eq.(%
\ref{a}) with eq.(\ref{aa}) 
\begin{equation}
\mathop{\rm Im}%
\mu =\frac{\hbar C^2}me^{-W/\hbar },
\end{equation}
which results in the decay rate 
\begin{equation}
\Gamma =\frac 12\frac 2\hbar 
\mathop{\rm Im}%
\mu =\frac{C^2}me^{-W/\hbar }
\end{equation}
where the factor $1/2$ comes from the analytical continuation. Physically,
this results from the assumption in the decay problem (and not in the MQC
problem) that the wave that has tunneled will never return. Mathematically,
it is due to the fact that the deformed contour runs from $0$ to $i\infty $,
and not from $-i\infty $ to $i\infty $\cite{Coleman}.

The constant $C$ can be determined from the normalization of the wave
function in the classically accessible region, which is connected with those
in the barrier region through\cite{Landau} 
\begin{equation}
\frac C{\sqrt{\left| p\right| }}\exp \left( -\frac 1\hbar \int_{r_1}^rpdr%
\right) \rightarrow \frac{2C}{\sqrt{\left| p\right| }}\cos \left( -\frac 1%
\hbar \int_{r_1}^rpdr-\frac \pi 4\right) .
\end{equation}
We restrict the integration in the classically accessible region, i.e., in
the potential well, $r<r_1$, since outside of this range $\psi $ decreases
exponentially. Because the argument of the cosine in the wave function is a
rapidly varying function, we can, with sufficient accuracy, replace the
squared cosine by its mean value 1/2. This gives 
\begin{equation}
C^2=\frac{m\omega }{2\pi }.
\end{equation}
Inserting this into the decay rate we have 
\begin{equation}
\Gamma =\frac \omega {2\pi }\exp \left[ -\frac W\hbar \right],   \label{gama}
\end{equation}
where $\omega $ is the frequency of the classical periodic motion 
\begin{equation}
\omega (\mu )=\frac{2\pi }{2m\int \frac{dx}p}=\frac \pi {\sqrt{2m}%
\int_0^{r_1}\frac{dr}{\sqrt{\mu -V(r)}}}  \label{omega}
\end{equation}
and can be calculated as 
\begin{equation}
\omega (\mu )=\omega _c\frac \pi 2\left[ \left( \frac 1{r_2}+Br_2\right)
K(k^{\prime })-Br_2E(k^{\prime })\right] ^{-1}
\end{equation}
with the complementary modulus $k^{\prime }=\sqrt{1-k^2}$. It must be
recalled that the frequency $\omega $ is in general different for different
levels, being a function of the chemical potential. We find that our
expression for the decay rate eq.(\ref{gama}) is more accurate than that of
Refs.\cite{Usama1,Usama2}, i.e., in the prefactor a chemical
potential dependent frequency replaces the constant attempt frequency $%
\omega _0$. In Figure 2 we show the dependence of this frequency on the
chemical potential; it decreases from $\omega _0$ as the chemical potential
increases from 0. This factor suppresses the tunneling rate greatly when the
chemical potential approaches the barrier top as shown in the figure, which
would be expected to increase the lifetime of the skyrmion.

\section{Numerical Results for Skyrmions}

\begin{figure*}
\includegraphics{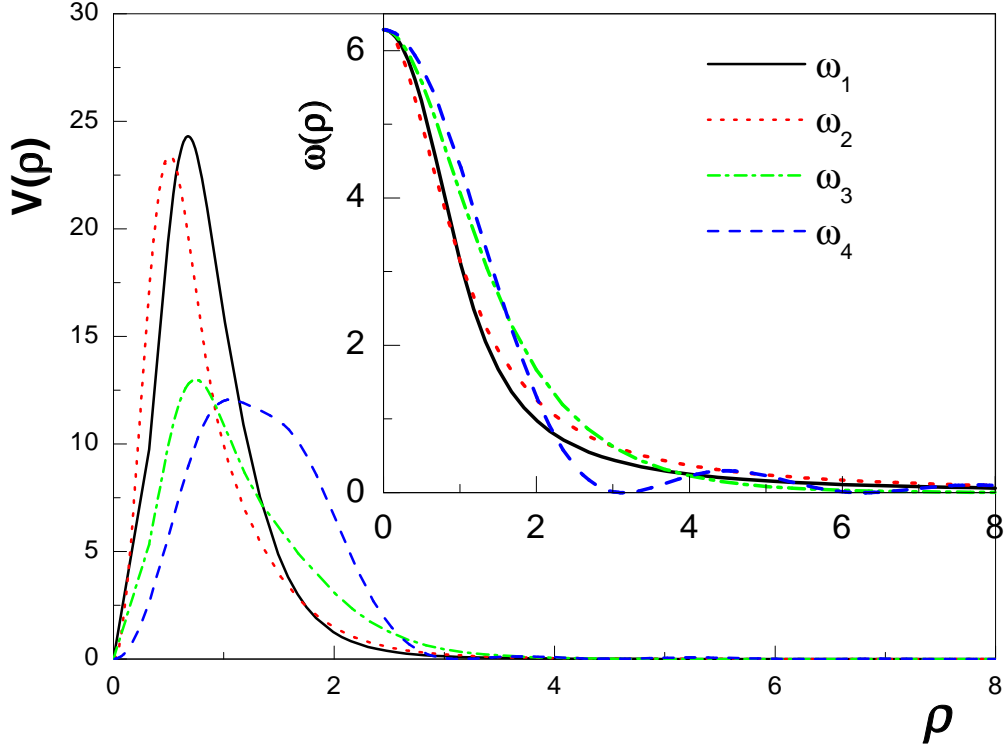}
\caption{Ansatze for $\omega (\rho )$ and the corresponding potential
barriers $V(\rho )$. The potential and the radius are again in units of 
$\hbar^2/2 m \xi^2$ and $\xi$, respectively.}
\end{figure*}

The skyrmions in a ferromagnetic condensate are energetically unstable as
shown in ref. \cite{Usama2}. The time scale on which the skyrmion shrinks
may be evaluated for two cases: For a large skyrmion, its size decreases at a
rate $\Gamma _{large}\approx 18\sec ^{-1}\xi /\lambda $ for $^{87}Rb$
spin-1/2 condensate of central density $10^{11}cm^{-3}$ and realistic
experimental conditions. For skyrmions with sizes of the order or less than
the correlation length $\xi $, the shrinking rate is determined by the
tunneling rate from the core of the skyrmion to the outer region. In previous studies
 \cite{Usama1,Usama2} the authors estimated roughly the lifetime of this small skyrmion due
to the tunneling process employing a WKB expression for the tunneling rate.
In this section we thus reconsider the lifetime using the result derived
above, with the modification originating from the prefactor included.

With an ansatz for $\omega (\rho )$ the problem is simplified to a nonlinear
Schr\"{o}dinger equation with the external potential of the off-centered
form. For a different functional behavior of $\omega (\rho )$ it turns out
that the effective potential will not be very different, as long as the
ansatz satisfies the boundary conditions $\omega (0)=2\pi $ and $\lim_{\rho
\rightarrow \infty }\omega (\rho )=0$ and falls off monotonically. 
Here we compute the decay rates of different skyrmion textures with the same size,
by taking into account two ansatze which were proposed in Ref. \cite{Usama2} as 
trial functions for simplifying the pair of nonlinear and coupled equations
(eq. (8) and (9) in ref.\cite{Usama2}). Namely 
\begin{eqnarray}
\omega _1(\rho ) &=&4\cot ^{-1}[(\rho /\lambda )^2], \\
\omega _2(\rho ) &=&\frac{2\pi }{1+(\rho /\lambda )^2},
\end{eqnarray}
where $\rho =r/\xi $ and $\lambda $ corresponds to the size of the skyrmion
and is also given in units of $\xi $. Considering the large distance behavior
of the coupling equations for $n(\rho)$ and $\omega(\rho)$, we see that for large 
skyrmions, the density fluctuations scale as $1/\rho^2$, and so should the
ansatze for $\omega(\rho)$. This is the reason why we would arrive at a non-physical 
result for a seemingly reasonable ansatz $\omega _3(\rho ) = 2 \pi \rm{sech}(\rho /\lambda )$.
We also check that non-monotonic behavior, i.e., oscillations in the 
falling of $\omega(\rho)$, for example, $\omega _4(\rho )=2\pi \left[ 
\frac{\sin (\rho /\lambda )}{\rho /\lambda }\right] ^2$, will inevitably
lead to singularities in the density profile, though the effective potential
holds an off-centered form. We show in Figure 3 these ansatze for $\omega (\rho )$ 
and their corresponding effective potentials $V(\rho )$.

One important parameter we should determine is the chemical potential of the
core atoms because we should know at which level the atom will tunnel out.
In principle one should solve the two-coupled nonlinear differential
equations and derive the density profile and spinor(or the function $\omega (\rho )$). 
As mentioned already above, we employ alternatively a simple approach, i.e., by introducing the
ansatz for $\omega (\rho )$. Then from the resulting density distribution we
calculate the energy for a particular value of $\lambda ,$ then the core
chemical potential can be calculated numerically by differentiating the
energy with respect to the number of core atoms. Performing the calculation
within a Thomas-Fermi approximation, which means in the expression for the
energy we neglect the kinetic energy term , we finally obtain the chemical
potential of the core atom $\mu $ for different values of $\lambda $ and
ansatz.

\begin{figure*}
\includegraphics{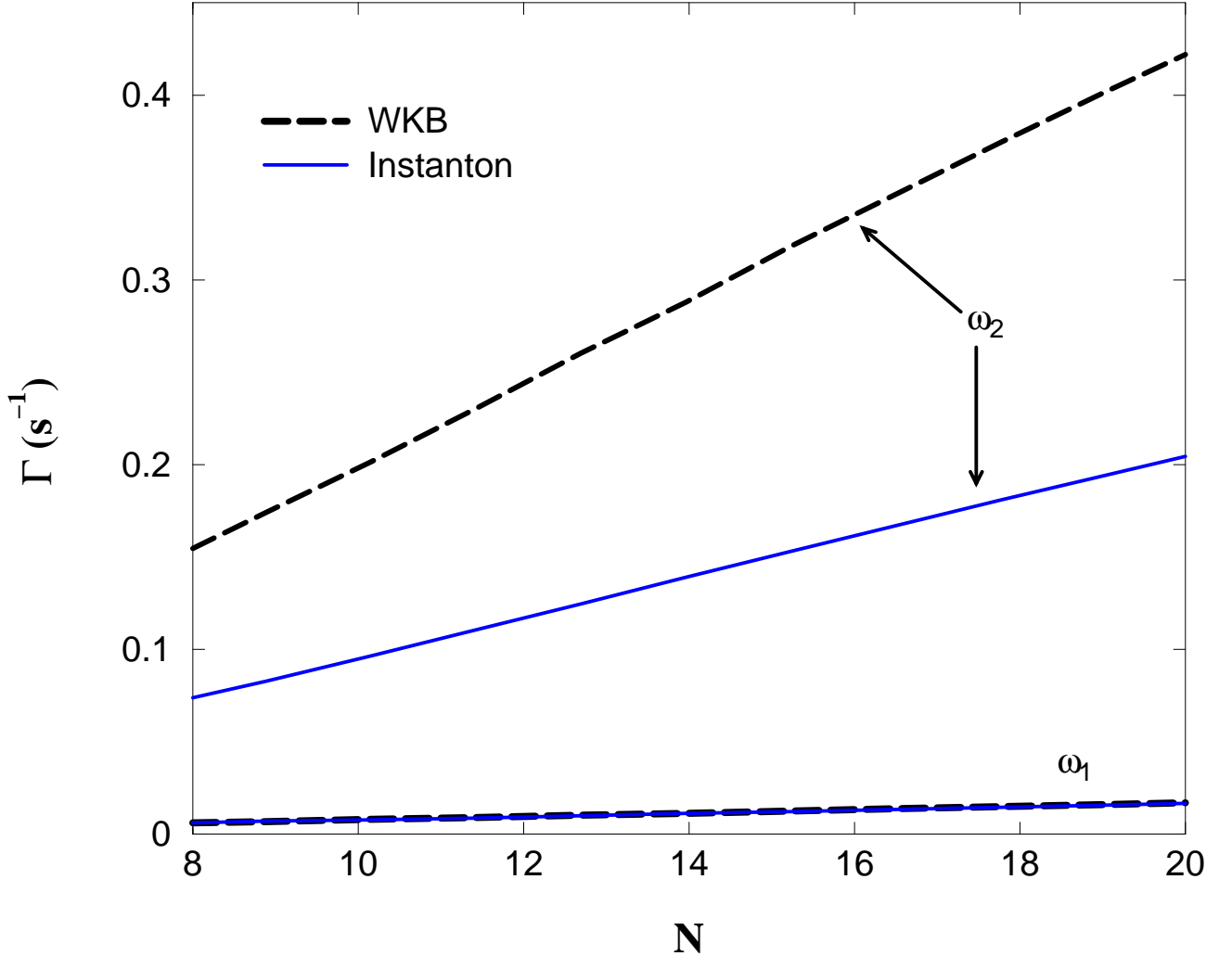}
\caption{The shrinking rate of skyrmions as a function of the number
of core atoms for $\omega_1$ and $\omega_2$. The calculation was performed 
for a $^{87}Rb$ spin-1/2 condensate with a scattering length of $a=5.4nm$.
The dashed lines are the WKB calculations in Ref. \cite{Usama1}, while the solid
lines show our results from the periodic instanton method.}
\end{figure*}

For the ansatze $\omega_1$ and $\omega_2$ we calculate the corresponding chemical potential for 
$\lambda =\xi $. The shrinking rates of the corresponding skyrmions are
calculated according to our decay rate expression eq. (\ref{gama}) with the
action given by eq.(\ref{w1}) and the prefactor given by eq.(\ref{omega}).
Figure 4 gives the tunneling rates as a function of the number of core
atoms. The calculation was performed for a $^{87}Rb$ spin-1/2 condensate
with a scattering length of $a=5.4nm$. 

We observe that \cite{Usama3} the correction resulting from the accurate prefactor $\omega(\mu)$ 
for $\omega_1$ is minor but significant for $\omega_2$. In Figure 2 we could generally
take the range of the chemical potential from $0$ to $V_m$. Unlike the situation
in a harmonic trap where the chemical potential $\mu$ increases with the number of
condensed atoms as $~N^{2/5}$ in the Thomas-Fermi approximation, in our case 
$\mu$ decreases with $N$ instead. This is because of the fact that the trap 
frequency (for $\omega_1$) $\omega_0$ is inversely proportional to the equilibrium 
skyrmion width $\lambda_0$, which in turn increases with N(apparently faster than $N^{1/5}$). 
Here $\lambda_0$ is determined from minimizing the total energy taking into account the outer
region of the skyrmion. This restricts us to a special domain of
$\mu$. For numbers of core atoms ranging from $1$ to $20$, $\mu/V_m$ 
ranges roughly from $0.16$ to $0.04$. In this interval 
$\omega(\mu)/\omega_0$ starts from $0.985$ for $N_{core}=1$
and ends at $0.998$ for $N_{core}=20$, which are almost indistinguishable in Figure 4. 
However, for ansatz $\omega_2$ the chemical potential (in units of 
$\hbar^2/2m\xi^2$) starts at $42$ for one core atom and
ends at $9.3$ for $20$ core atoms. The corresponding correction is shown in Figure 4.

Till now there is still no clear experimental evidence for the skyrmions in the condensate.
From the above calculations we see that the result for the decay rate depends crucially on the
detailed form of the ansatz $\omega$.
It remains a challenging task to solve the coupled nonlinear equations numerically,
and to compare the results with those above.

\section{Conclusion}

We present here an accurate calculation of the tunneling rate for a class of
off-centered potentials with a periodic instanton method. Apart from its
application to the study of the stability of the skyrmion excitation in the
two-component ferromagnetic condensate, the bounce for the off-centered
potential barrier is itself a novel configuration from the viewpoint of the
scalar field theory. The exact prefactor of the decay rate has been calculated
and we found it depends on the chemical potential at the level of the atoms
tunneling to the outer region. This modifies the result for the rough estimate
of the lifetime by a constant attempt frequency $\omega _0$. One can easily
find some similar off-centered potentials in other topological excitations
such as vortices, monopoles, etc. Our periodic instanton formalism can be
extended to the investigation of the lifetime and tunneling behavior in
these systems. Further studies should include the properties of the
quantum-classical transition of the decay rate when the chemical potential
increases and surpasses the barrier height.

\acknowledgments  We thank Usama Al Khawaja and Henk Stoof for their help
in numerical simulation, especially for providing us Figure 4. It is a great 
pleasure to thank J.-Q. Liang and Yaping Yang for useful discussions. Y.Z.
acknowledges support by an Alexander von Humboldt Foundation Fellowship.
This research was supported in part by NSFC of China under grant No.
10175039 and No. 10075032.

\end{document}